\documentclass[a4paper,11pt]{article}
\pdfoutput=1 

\usepackage{jheppub} 

\usepackage[T1]{fontenc} 

\newcommand{\cN}{\mathcal{N}}

\newcommand{\ri}{{\rm i}}
\newcommand{\p}{{\partial}}
\newcommand{\s}{{\sigma}}
\renewcommand{\u}[1]{\underline{#1}}

\title{\rm \bf \Huge Internal Supersymmetry and Small-field Goldstini}

\author[a]{Diederik Roest,}
\author[a]{Pelle Werkman,}
\author[b]{Yusuke Yamada.}

\affiliation[a]{Van Swinderen Institute for Particle Physics and Gravity, University of Groningen, \\ Nijenborgh 4, 9747 AG Groningen, The Netherlands}
\affiliation[b]{Stanford Institute for Theoretical Physics and Department of Physics, Stanford University,\\382 Via Pueblo Mall
Stanford CA 94305, USA}

\emailAdd{d.roest@rug.nl}
\emailAdd{p.j.werkman@rug.nl}
\emailAdd{yusukeyy@stanford.edu}

\abstract{The dynamics of the Goldstino mode of spontaneously broken supersymmetry is universal, being fully determined by the non-linearly realized symmetry. We investigate the small-field limit of this theory. This model non-linearly realizes an alternative supersymmetry algebra with vanishing anti-commutators between the fermionic generators, much like an internal supersymmetry. This Goldstino theory is akin to the Galilean scalar field theory that arises as the small-field limit of Dirac-Born-Infeld theory and non-linearly realizes the Galilean symmetry. Indeed, the small-field Goldstino is the partner of a complex Galilean scalar field under conventional supersymmetry. We close with the generalization to extended internal supersymmetry and a discussion of its higher-dimensional origin.}

\begin{document} 
\maketitle
\flushbottom
\section{Introduction}

Symmetries in their various guises form the cornerstone of modern physics. Of particular interest is the case of spontaneously broken symmetries, either of internal \cite{CWZ, CCWZ} or space-time nature \cite{Volkov}. Both cases lead to Goldstone modes that transform in a non-linear representation. The resulting Goldstone dynamics is characterized by a small number of coefficients and therefore has clear signatures, including masslessness at quadratic order and special soft limits at higher orders \cite{Cheung, CheungPeriodic, Padilla}. 

A beautiful example is provided by a scalar in $D=4$ Poincar\'e invariant field theories. The scalar can arise as a Goldstone boson from various extensions of the space-time symmetry, including the well-known possibilities of $D=4$ conformal and $D=5$ Poincar\'e. The allowed interactions for these scalar fields have been constructed from probe brane constructions  \cite{Reunited, Hinterbichler}. For Poincar\'e, this includes the Dirac-Born-Infeld (DBI) action and its higher-derivative versions. These are invariant under 
 \begin{align}
   \delta \phi = c + b_\mu (x^\mu + \phi \partial^\mu \phi) \,, \label{Poinc-tr}
  \end{align}
corresponding to the translation $P_5$ and the rotation $M_{\mu 5}$ of the higher-dimensional algebra. 

The third possibility arises as a singular contraction of the Poincar\'e (or conformal) algebra, defined as
 \begin{align}
  P_5 \rightarrow \omega P_5 \,, \quad  M_{\mu 5} \rightarrow \omega M_{\mu 5} \,, \quad {\rm with~} \omega \rightarrow \infty \,. \label{Galilean-limit}
 \end{align}
It preserves the $D=4$ Poincar\'e part, and in addition has
 \begin{align}
  & [P_5, M_{\mu 5} ] = 0 \,, \quad [P_\mu, M_{\nu 5} ] = i \eta_{\mu \nu} P_5 \,, \notag \\ 
  & [ M_{\mu \nu}, M_{\rho 5} ] = i (\eta_{\mu \rho} M_{\nu 5} - \eta_{\nu \rho} M_{\mu 5}) \,.  \label{Gal-algebra}
 \end{align}
The resulting {\it Galilean algebra} differs from the $D=5$ Poincar\'e algebra only in the first commutator being zero. Moreover, it is a non-trivial extension of the space-time group due to the non-vanishing of the second commutator; the final commutator just expresses the Lorentz nature of the additional vector generator. 

The same limit can be performed on the field theory side. DBI can be written as a coset in terms of the combination $\phi P_5$ and is invariant under \eqref{Galilean-limit} with
 \begin{align} 
\phi \rightarrow \phi / \omega \,,  \quad {\rm with~} \omega \rightarrow \infty \,. \label{small-field}
\end{align}
In this singular, small-field limit, one obtains an inequivalent coset that is based on the contracted, Galilean algebra. The non-linear transformation becomes \cite{Reunited}
 \begin{align}
    \delta \phi = c + b_\mu x^\mu \,, \label{Gal-tr}
  \end{align}
and hence is field-independent. {The leading interactions for this theory, referred to as \emph{Galileons}, are Wess-Zumino terms  \cite{GalileonWessZumino}.} The cubic one describes the brane bending mode of the DGP model \cite{DGP}, and more generally these interactions describe models of massive gravity in their decoupling limits~\cite{ReviewMassiveGravity}. Moreover, their covariantized version \cite{CovGalileons} has been used to describe late-time acceleration in cosmology; however, this particular setup may be ruled out by recent observations \cite{Peirone}. 

The above three theories with  enhanced  (Poincar\'e, conformal or Galilean) symmetry are unique in having a special soft limit of the amplitudes \cite{Cheung, CheungPeriodic,Padilla}; this is intimately tied to the extended and non-linear symmetries of these particular theories.

One might wonder whether there is a similar pattern on the fermionic side, with Grassmannian extensions $Q_\alpha$ of the Poincar\'e algebra. Indeed such a construction is possible for $\cN = 1$ supersymmetry, leading to the Volkov-Akulov (VA) Goldstino with transformation \cite{Volkov-Akulov,IvanovKapustnikov1,IvanovKapustnikov2}
 \begin{align}
  \delta \lambda^\alpha = \epsilon^\alpha - i (\lambda \sigma^\mu \bar \epsilon - \epsilon \sigma^\mu \bar \lambda ) \partial_\mu \lambda^\alpha \,. \label{VA-tr}
 \end{align}
Similar to the bosonic construction, this theory can be interpreted as a three-brane in superspace \cite{Kallosh} and has an enhanced soft limit as compared to generic interacting fermion theories \cite{Kallosh-Karlsson}.

Could there be different supersymmetry algebras\footnote{This might appear to be ruled out by the superalgebra classification of \cite{Haag}. However, similar to the Coleman-Mandula theorem on extensions of space-time symmetry \cite{Coleman-Mandula}, this only concerns superalgebras with linear representations.} leading to inequivalent Goldstino dynamics? We will demonstrate that this is indeed the case and construct an alternative, which we will refer to as internal supersymmetry. It acts on the corresponding Goldstino mode as a constant shift, allowing for specific interactions only (see Section 2). Such a symmetry appears in the limit where the corresponding Goldstino (super)field fluctuation is very small: the Goldstino describes small fluctuations of a supersymmetric brane, which (partially) breaks supersymmetry as well as Poincar\'e symmetry in four (or higher) dimensions.

Under conventional linear supersymmetry, the non-linear transformation laws of fermions and bosons are related. Indeed this happens for the field-dependent transformations \eqref{Poinc-tr} and \eqref{VA-tr}, which are invariances of $\cN = 1$ super-DBI theory constructed in \cite{Bagger:1996wp,Bagger:1997pi,RocekTseytlin,GonzalezParkRocek}. Phrased differently, this theory has non-linearly realized Poincar\'e and supersymmetry generators that commute with linear $D=4$ super-Poincar\'e  into $D=6$ super-Poincar\'e, and the non-linearly realized generators commute into each other under the linearly realized supersymmetry. A natural question regards the non-linear symmetries of this theory in the small-field limit. We will demonstrate that these become the Galilean symmetry and internal supersymmetry, which again are related under linear supersymmetry (see Section 3).  

We further generalize our construction by considering the appropriate non-relativistic contraction of the $D=10$ supersymmetry algebra instead. This leads to a relativistic $D=4$ theory with linearly realized minimal supersymmetry, as well as non-linearly realized $\mathcal N$-extended internal supersymmetry (see Section 4).

\section{The internal supersymmetry algebra and Goldstino}

To demonstrate how one can obtain the alternative supersymmetry algebra by means of a simple contraction, we start with the $\cN = 1$ super-Poincar\'e algebra in\footnote{For definiteness of fermionic conventions, we will focus on $D=4$, but much of our discussion carries over to other dimensions.} $D=4$, with fermionic generators $Q_\alpha$ subject to
\begin{align}
&\{Q_{\alpha},\overline{Q}_{\dot\beta}\}=-2\ri \sigma^\mu_{\alpha\dot\beta}P_\mu \,, \quad [M_{\mu\nu},Q_{\alpha}]=(\sigma_{\mu\nu}Q)_{\alpha} \,.
\end{align}
Under the $\dot{\text{I}}$n\"on\"u-Wigner contraction 
 \begin{align}
  Q_\alpha \to \omega S_\alpha \,,  \quad {\rm with~} \omega \rightarrow \infty \,, \label{int-limit}
  \end{align}
where we have relabelled the supercharge generator as $S$ to emphasize its physical difference, this algebra becomes
\begin{align}
&\{S_{\alpha},\overline{S}_{\dot\beta}\}= 0 \,, \quad [M_{\mu\nu},S_{\alpha}]=(\sigma_{\mu\nu}S)_{\alpha} \,. \label{internal-SUSY}
\end{align}
Finally, the supersymmetry generators have a non-trivial weight under a $U(1)$ R-symmetry.

Note that the hallmark of the supersymmetric extension of the Poincar\'e space-time symmetry group, i.e.~that supercharges anti-commute into translations, has disappeared in this limit. The only non-trivial commutator is with Lorentz generators, reflecting the fact that $S$ transforms as spin-1/2. This fermionic extension of Poincar\'e is  akin to an internal symmetry, and we will refer to it as {\it internal supersymmetry}. 

The above algebra contraction can be seen as the small-field limit of the $\cN = 1$ super-Poincar\'e theory. Again the coset combination  $\lambda^\alpha Q_\alpha$ is invariant under the algebra contraction together with the fermion small-field limit
 \begin{align}
  \lambda^\alpha \to \lambda^\alpha / \omega \,, \quad {\rm with~} \omega \rightarrow \infty  \,.
 \end{align}
In this limit, the original VA transformation of the Goldstino $\lambda^\alpha$ \eqref{VA-tr} yields a simple fermionic shift,
\begin{align}
\delta \lambda^\alpha = \epsilon^\alpha \,, \label{ferm-tr}
\end{align}
where $\epsilon^\alpha$ is a constant parameter. 

Note the clear analogy between the field-dependent transformations of DBI \eqref{Poinc-tr} and VA \eqref{VA-tr}, and their small-field limits \eqref{Gal-tr} and \eqref{ferm-tr}: in both cases one loses the field dependence in the small-field limit, corresponding to the vanishing of the crucial first commutators in \eqref{Gal-algebra} and \eqref{internal-SUSY}. One would expect the contracted algebras to only have non-linear representation, and internal supersymmetry therefore cannot be restored as a linear transformation above some energy scale.
Indeed, this theory has been shown not to admit unitary UV completions \cite{Bellazzini:2016}, similar to Galileons; instead, it could provide an alternative realization of fermion compositeness with specific LHC signatures, as discussed in \cite{Liu, Bellazzini:2017}.

Turning to invariants, the lowest order invariant (up to a total derivative) is the simple fermionic kinetic term, 
\begin{align}
i \bar \lambda \sigma^\mu \partial_\mu \lambda \,, \label{Dirac}
\end{align}
which arises as the small field limit of the VA invariant~\cite{Volkov-Akulov,IvanovKapustnikov1,IvanovKapustnikov2}. While the latter includes interactions, the theory becomes free in the small-field limit\footnote{An analogous story can be found on the bosonic side, where the lowest-order invariant becomes the free kinetic theory in the Galilean limit, while higher-order invariants introduce cubic and quartic interactions.}. 

We can see that further Wess-Zumino terms do not exist in $D=4$ by using the coset formalism going back to \cite{CWZ, CCWZ}. The invariant 1-forms appearing in the decomposition of the Maurer-Cartan form are the following: 
\begin{align}
\omega_P^\mu = dx^\mu,\quad \omega_S^\alpha = d\lambda^\alpha \,,
\end{align}
where $\lambda$ is Majorana. The Wess-Zumino terms are obtained by wedging these 1-forms together to a 5-form living in a space in which $\lambda^\alpha$ is promoted to a coordinate. This form must then be pulled back to the four-dimensional space defined by $\lambda = \lambda(x)$. Wedging together the 1-forms immediately implies anti-symmetrization of all derivatives of $\lambda$ appearing in a Wess-Zumino term. The possibilities for quartic interaction terms are then highly limited by Lorentz invariance. All possible terms are of the following form:
\begin{align}
(\bar{\lambda}\gamma\partial \lambda) (\partial \bar{\lambda}\gamma\partial\lambda) \,,
\end{align}
where $\gamma$ denotes a gamma matrix of some rank. Both gamma matrices must be rank one or two, since other choices are immediately vanishing or total derivative due to Majorana flip relations. The only possible terms are then:
\begin{align}
(\bar{\lambda}\gamma^{\mu_1} \partial_{[\nu_1} \lambda) (\partial_{\nu_2} \bar{\lambda}\gamma^{\mu_2 \mu_3}\partial_{\nu_3]} \lambda) \,.
\end{align}  
with some product of Kronecker deltas and Levi-Civita tensors contracting the indices. However, each term is trivial due to a Fierz identity. It therefore appears that, at least in $D=4$, the situation for fermions is akin to that of vectors, for which a no-go theorem for Galileon-like interactions was proven in \cite{Deffayet}. It would be interesting to investigate whether non-trivial Wess-Zumino terms exist in dimensions higher than 4, where the structure of Fierz identities and Majorana flips is different and Wess-Zumino terms beyond quartic order in fermions can exist.

In the absence of Wess-Zumino terms, the fermion is derivatively coupled in all  invariants, which implies that they give rise to second-order field equations and Ostrogradsky instabilities. In some cases, however, a supersymmetric coupling to a healthy bosonic sector can remove the instability, as the example of the next section demonstrates. 

\section{Adding linear supersymmetry and Galileons}

Our second goal will be to show that internal supersymmetry can be combined with a linearly realized supersymmetry of conventional nature, under which it is the natural partner of Galilean transformations. A useful starting point will be the minimal supersymmetry algebra in $D=6$, which has eight supercharges and $SU(2)$ Majorana-Weyl spinors (see e.g.~\cite{Abe:2015bqa}). Rewriting this algebra in terms of $D=4$ Weyl spinors via 
 \begin{align}
   Q_{1 \u\alpha}=(Q_\alpha, -\overline{S}^{\dot\alpha})^T \,, \quad Q_{2 \u\alpha}=(S_{\alpha},\overline{Q}^{\dot\alpha}) \,,
\end{align} 
where $\u\alpha$ is a 4-component spinor index of the $SU(2)$ Majorana-Weyl spinor, the anti-commutators of the supercharges read
\begin{align}
& \{Q_{\alpha},\overline{Q}_{\dot\beta}\}=\{S_{\alpha},\overline{S}_{\dot\beta}\}=-2\ri\sigma^\mu_{\alpha\dot\beta}P_\mu, \notag \\
& \{Q_{\alpha},S_{\beta}\}=2 \epsilon_{\alpha\beta} P_z \,, \quad \{\overline{Q}^{\dot\alpha},\overline{S}^{\dot\beta}\}=-2\epsilon^{\dot\alpha\dot\beta} P_{\overline z},
\end{align}
where $z = \tfrac12( x^4 - \ri x^5)$ and $P_z=P_4+\ri P_5$ in our conventions. This is the $\cN =2$-extended supersymmetry algebra in $D=4$, with $U(2)$ R-symmetry group and $SO(1,5)$ automorphisms inherited from its six-dimensional origin.

We now consider the Galilean rescaling \eqref{Galilean-limit} of this algebra. Importantly, this limit would be incompatible without scaling the fermions as well: the anti-commutator would become singular. Therefore there is no $\cN = 2$ extension of the Galilean algebra with the usual supersymmetry. Instead, the interesting and consistent option is to rescale one component of the $U(2)$ doublet, which we take to be $S$ without loss of generality:
 \begin{align}
  Q_\alpha \rightarrow Q_\alpha \,, \quad S_\alpha \rightarrow \omega S_\alpha \,, \quad {\rm with~} \omega \rightarrow \infty \,,
 \end{align}
in addition to the rescalings \eqref{Galilean-limit} of the bosonic generators $P_z$ and $M_{\mu z}$. In addition, the off-diagonal generators of $SU(2)$, denoted by $R$ and its conjugate, both scale as 
 \begin{align}
    R\to \omega R \,, \qquad \bar R \to \omega \bar R \,,  \quad {\rm with~} \omega \rightarrow \infty \,, \label{fund-rescaling}
\end{align}
for consistency.

After the contraction, the anti-commutators between the different supersymmetry generators are
\begin{align}
& \{Q_{\alpha},\overline{Q}_{\dot\beta}\}=-2\ri \sigma^\mu_{\alpha\dot\beta}P_\mu \,, \quad  \{S_{\alpha},\overline{S}_{\dot\beta}\}=0 \,, \notag \\
& \{Q_{\alpha},S_{\beta}\}=-2\epsilon_{\alpha\beta} P_z \,,  \quad
 \{\overline{Q}^{\dot\alpha},\overline{S}^{\dot\beta}\}=2 \epsilon^{\dot\alpha\dot\beta} P_{\bar z} \,,
\end{align}
where only the second anti-commutator differs from the original algebra. Secondly, Lorentz boosts in the contracted directions satisfy  
\begin{align}\label{eq:contractedcommutators}
  [ M_{\mu \bar z} ,Q_\alpha] = \tfrac{1}{2} \ri (\sigma_\mu\overline{S})_{\alpha} , \quad [M_{\mu \bar z} ,S_\alpha]=0 \,.
 \end{align}
The final set of rescaled generators has commutators
\begin{equation}
[R,Q_\alpha]=S_\alpha\,,\quad [R, S_\alpha]=0 \,,
\end{equation}
Finally, this algebra inherits two copies of $U(1)$ from its original R-symmetry group.

This algebra can be seen as an extension of the usual $\cN = 1$ super-Poincare algebra with the rescaled Galilean-like generators, all of which are realized non-linearly. Note that they form a sequence under linear supersymmetry: $M_{\mu z}$ and $R$ transform into $S$ under $Q$, which in turn is transformed to\footnote{A possible further extension of this sequence would be the special Galileon symmetry \cite{special-Gal} of a real scalar field; however, no such symmetry is known for the complex case.} $P_z$:
 \begin{align}
   Q_\alpha: \quad (M_{\mu z},R) \rightarrow S_\alpha \rightarrow P_z \rightarrow 0 \,, \label{Gal-gen}
 \end{align}
Moreover, this extension is not fully internal due to the non-trivial commutator of $M_{\mu z}$ with translations. 

Turning to realizations, a natural formulation of the non-linear symmetries of this theory presents itself in terms of a superfield of the linear supersymmetry. Consider a chiral superfield $\Phi$ given by 
 \begin{align}
 \Phi = \phi + \lambda \theta + F \theta  \theta \,,
 \end{align}
consisting of a complex scalar field $\phi$, a fermion $\lambda$ as well as an auxiliary field $F$. Under linear supersymmetry, these components transform as 
 \begin{align}
   \delta \phi = \bar \epsilon \lambda \,, \quad 
   \delta \lambda =\bar\epsilon\sigma^\mu\partial_\mu \phi+F\epsilon\, \quad 
    \delta F = \bar \epsilon \sigma^\mu \partial_\mu \lambda \,. 
  \end{align}
Moreover, the Galileon-like generators \eqref{Gal-gen} act on this superfield as
\begin{equation}
	\Phi \rightarrow \Phi + c +\theta \eta+ b_\mu (x^\mu + i\theta \sigma^\mu\bar{\theta} ) + f \theta \theta \,,
\end{equation}
or equivalently,
\begin{equation}
\phi\to\phi+c+b_\mu x^{\mu}\, ,\quad \lambda_\alpha\to \lambda_\alpha+\eta_\alpha\,,\quad  F\to F + f \,,
\end{equation}
with constant parameters, corresponding to the generators $P_z$, $M_{\mu z}$, $S_\alpha$ and $R$, respectively. Importantly, these all preserve the chiral nature of the superfield. 

Finally, the invariant Lagrangians can be written in terms of superfields. At lowest order, the ordinary kinetic terms follow from the usual superspace expression $\Phi\overline{\Phi}$, which includes the Dirac action \eqref{Dirac} for the fermionic component. The first interactions, at the quartic level, can be classified according to which of the generators in the sequence \eqref{Gal-gen} are realized. Its smallest part is the bosonic shift symmetry with the invariant $D \Phi D \Phi \bar D \bar \Phi \bar D \bar \Phi$, including {the purely bosonic term} $(\partial \phi)^4$ at mass dimension-8, see e.g.~\cite{Khoury}. The entire sequence is realized by\footnote{At the cubic level in $\Phi$, there are no Wess-Zumino terms of \eqref{Gal-gen} which contain a Galileon term. See \cite{Lehners2}.} \cite{Farakos}:
\begin{align}
 \Phi(\overline{D}_{\dot\alpha}\partial_\mu\overline{\Phi}\bar{\sigma}_{\nu}^{\dot\alpha\alpha}D_{\alpha}\partial_\rho\Phi)\epsilon^{\mu\nu\rho\sigma}\partial_{\sigma}\overline{\Phi} \,. \label{quartic-invariant}
\end{align} 
This is the first non-trivial Wess-Zumino term for the sequence \eqref{Gal-gen}. This interaction plays a similar role to the purely bosonic Galileon, as it describes the fluctuations of a brane in superspace, after taking the small-field limit of the bending modes and the Goldstino. When truncated to the complex scalar\footnote{An analogous construction for a $D=3$ superfield containing a real scalar can be found in \cite{Queiruga}.}, it is proportional to the usual quartic Galileon at mass dimension-10:
 \begin{align}
   \phi (\partial_{\mu_1} \partial^{[\mu_1} \phi)  (\partial_{\mu_2} \partial^{\mu_2} \bar \phi)  (\partial_{\mu_3} \partial^{\mu_3]} \bar \phi) \,.
 \end{align}
At the same order, the  fermionic contribution reads
 \begin{align}
  -4i \p_\mu  \lambda \s^\tau  \p_\tau  \bar{\lambda} \p_\s  \bar{\lambda} \bar{\s}_\nu \p_\rho  \lambda \epsilon^{\mu \nu \rho \s } + \notag -2i \p_\mu  \bar{\lambda} \bar{\s}_\nu \s^\kappa  \p_\s  \bar{\lambda }  \epsilon^{\mu \nu \rho \s }  \p_\kappa  \lambda \p _\rho  \lambda \,.
  \end{align}
While these terms are manifestly invariant under internal supersymmetry, this is not the case for the mixed scalar-fermion terms, which can be found in \cite{Farakos}; however, one can check that these always multiply total derivatives and hence do not affect the field equations. 

By construction, the bosonic field equations have the Galileon structure and hence do not propagate any ghosts. The fermionic sector, however, necessarily has second-order field equations and hence seems to propagate a ghost. This seems a paradoxical conclusion, as the theory is also linearly supersymmetric and hence cannot have purely fermionic ghosts. We believe this to be resolved by the coupling between the bosons and fermions of this theory, along the lines of \cite{Langlois, Klein, Kimura}.

\section{Extended internal supersymmetry and Goldstini}

Finally, we will demonstrate that it is possible to have $\cN$-extended internal supersymmetry, in addition to linearly realized supersymmetry. Instead of taking  $D=6$   as a starting point, one can perform a similar analysis for $D=10$ minimal supersymmetry. {Its supersymmetry generator is a Majorana-Weyl spinor that decomposes into four 4D Majorana spinors $Q_i$ with $i=1,\ldots,4$. As in the 6D case, the rotation in the extra dimensions becomes part of the R-symmetry, spanning the adjoint $\bf 15$ of $SO(6) \simeq SU(4)$. The supersymmetry generators $Q_i$ transform in the fundamental of $SU(4)$, while the translations $P_m$ as well as the Lorentz transformations $M_{\mu m}$ with $m=4,\ldots,9$ form a ${\bf 6}$ self-dual representation.  

In analogy with the 6D situation, the contraction of the algebra requires the decomposition of $SU(4)$ into $SU(3) \times U(1)$, with $\bf 4 \rightarrow \bf 3 + \bf 1$ and $\bf 6 \rightarrow \bf 3 + \bar {\bf 3}$ or in terms of indices $i = (I,4)$ and $m=(I, \bar I)$. We then perform the scaling 
\begin{align}
 & P_I \to \omega P_I \,, \quad M_{\mu I} \to \omega M_{\mu I}, \notag \\
& Q_{I \alpha} \to \omega S_{I \alpha}\,,  \qquad {\rm with~} \omega \rightarrow \infty \,,
\end{align}
where $I =1,2,3$. Similar to the previous section, this is not sufficient for a consistent contraction as the commutator between $Q \equiv Q_4$ and some of the $M_{mn}$  diverges. One needs to split up the R-symmetry generators as well, with $\bf 15 \rightarrow \bf 8 + \bf 3 + \bar {\bf 3} + \bf 1$. We denote the ${\bf 3}$ and $\bar{\bf 3}$ generators as $R_I$ and $\bar{R}_{\bar I}$, and rescale both similar to \eqref{fund-rescaling}.

The resulting algebra takes the following form. In addition to conventional supersymmetry with generators $Q$, it contains internal supersymmetry generators $S_{I \alpha}$ with
 \begin{align}
  & \{Q_{\alpha}, \overline{Q}_{\dot\beta}\}=-2\ri \sigma^\mu_{\alpha\dot\beta}P_\mu \,, \quad  \{S_{I \alpha}, \overline{S}_{\bar J \dot\beta}\}=0 \,, \notag \\
 & \{Q_{\alpha},S_{I \beta}\}=-2\epsilon_{\alpha\beta} P_I ,\quad \{\overline{Q}^{\dot\alpha},\overline{S}_{\bar I}^{\dot\beta} \}=2 \epsilon^{\dot\alpha\dot\beta} P_{\bar I} \,.
 \end{align}
Here we have defined $P_I = P_{2+2I} + \ri P_{3+2I}$, which together with $M_{\mu I}$ are the Galilean transformations that form the partners of internal supersymmetries: 
 \begin{align}
  [ M_{\mu \bar I}, Q _\alpha] = \tfrac{1}{2} \ri (\sigma^\mu\overline{S})_{\bar I \alpha}\,, \quad [M_{\mu \bar I}, S_{J \alpha}]=0 \,.
 \end{align}
Finally, one has
\begin{align}
[R_I, Q_\alpha]=S_{I\alpha}\, ,\quad [R_I, S_{J\alpha}]=0 \,,
\end{align}  
for the rescaled off-diagonal R-symmetries. Note that all rescaled, non-linearly realized generators transform in the fundamental of $SU(3)$, and form a sequence analogous to \eqref{Gal-gen} under conventional supersymmetry. In addition this algebra allows for a $U(1)$ acting on $Q$.

The superspace action \eqref{quartic-invariant} can easily be extended to 
\begin{align}
  \Phi^I(\overline{D}_{\dot\alpha}\partial_\mu\overline{\Phi}^{\bar{I}}\bar{\sigma}_{\nu}^{\dot\alpha\alpha}D_{\alpha}\partial_\rho\Phi^J)\epsilon^{\mu\nu\rho\sigma}\partial_{\sigma}\overline{\Phi}^{\bar J}  \,. \label{extended-quartic}
\end{align} 
This is the unique generalization of the quartic invariant for a single superfield that is compatible with the $U(1) \times U(\cN)$ symmetry. Its invariance under the non-linearly realized internal supersymmetry can be seen in a very analogous manner as the discussion in section III.

Despite the absence of linear supersymmetry representations without higher spins, there are analogous superalgebras in dimensions higher than ten, allowing for similar contractions. In this manner one can obtain the $\cN$-extended generalization of the $D=6$ and $D=10$ results for arbitrary $\cN$. Since the resulting algebras do not have any linear representations, and can be realized solely on spinors, this does not contradict the common statements that global supersymmetry has $\cN \leq 4$ in $D=4$.

\section{Conclusions}

We have discussed the existence of a fermionic symmetry akin to bosonic internal symmetries. Being spontaneously broken, its Goldstino mode transforms with a constant fermionic parameter, reminiscent of Goldstone bosons for internal symmetries. This fermionic extension of Poincar\'e is inequivalent to conventional supersymmetry and instead can be constructed from $\dot{\text{I}}$n\"on\"u-Wigner contractions of superalgebras, starting either in four or in higher dimensions. The latter case leads to linearly supersymmetric theories in which internal supersymmetry is the natural partner of the Galileon algebra.

Effective field theories have universal dynamics for the Goldstone modes, with particular signatures. In the case at hand, internal supersymmetry ensures the vanishing of its amplitudes in the soft limits; this is the fermionic analogon of the Adler zero, see the recent discussion \cite{Kallosh-NLS}. An analysis analogous to \cite{Cheung, CheungPeriodic} could provide a periodic table of fermion effective field theories with soft limits. 

We have demonstrated that no Wess-Zumino terms for internal supersymmetry exist in $D=4$ beyond the ordinary kinetic term. The fermion is therefore always derivatively coupled in absence of coupling to different fields. However, the supersymmetric Galileons of \cite{Farakos} provide an interaction which is nonetheless free of Ostrogradsky ghosts. It appears that the higher-dimensional origin of these constructions plays an important role in eliminating possible ghost degrees of freedom, as this particular realization of supersymmetric Galileon is shown here to satisfy a contracted higher-dimensional symmetry algebra. This connection was found long ago in the uncontracted case \cite{Hughes} and appears to be the same in the contracted case \cite{Lehners2, Farakos}.

All degrees of freedom considered in this paper are Goldstone modes, whose dynamics can be characterized in terms of a small number of coupling coefficients (in contrast to generic supersymmetry theories with e.g.~K\"{a}hler and superpotentials). Important questions that we leave for future work include the coupling to matter, e.g.~in $\cN = 1$ superspace. Another generalization involves other multiplets than the chiral superfield $\Phi$; for instance, it would be interesting to investigate the relation between the real linear superfield and the $D=5$ superalgebra. Because the real linear contains only a single scalar field, the $D=5$ translations and Galilean transformations can be realized on it while maintaining the constraints on the superfield. This might shed light on the higher-dimensional origin of supersymmetric theories based on a real Galileon scalar considered in \cite{KhouryLehnersOvrut} and/or the fermionic rescaling of \cite{Kamimura}. 

{The universal nature of Goldstone dynamics provides for a strong motivation to further elucidate these matters.}


  \section*{{Acknowledgments}} 
  
  We are grateful to F.~Farakos, R.~Kallosh, R.~Klein and D.~Stefanyszyn  for stimulating discussions, {and in particular S.~Garcia-Saenz for pointing out the Fierz identity for the putative quartic fermion interaction.} The work  of YY is supported by SITP and by the US National Science Foundation grant PHY-1720397. PW acknowledges the Dutch funding agency Netherlands Organisation for Scientific Research (NWO) for financial support. YY is grateful to the University of Groningen for the hospitality when this work was initiated.

\end{document}